\documentclass[draft]{agujournal2019}
\usepackage{apacite}
\usepackage{url}   
\usepackage[inline]{trackchanges} 
\usepackage{soul}
\usepackage{amsmath}
\usepackage{xr}
\draftfalse
\journalname{Geophysical Research Letters}

\begin{document}


\title{From spinning sea ice floes to ocean enstrophy spectra in the Marginal Ice Zone}  


\authors{M. Kim \affil{1}, G.E. Manucharyan \affil{2}, M. H. DiBenedetto \affil{3}, E. M. Buckley\affil{4}, D. M. Watkins \affil{1}, M. M. Wilhelmus \affil{1}}
\affiliation{1}{Center for Fluid Mechanics, School of Engineering, Brown University, Providence, RI 02912, USA}
\affiliation{2}{School of Oceanography, University of Washington, Seattle, WA, USA}
\affiliation{3}{Department of Mechanical \& Aerospace Engineering, Princeton University, Princeton, NJ 08544, USA}
\affiliation{4}{Department of Earth Science and Environmental Change, University of Illinois, Urbana-Champaign, IL 61280, USA}


\correspondingauthor{Monica Martinez Wilhelmus}{monica\_martinez\_wilhelmus@brown.edu}



\begin{keypoints}
\item By tracking ice floes and using them as natural tracers, we reveal the spectral characteristics of upper-ocean turbulence across the Arctic marginal ice zone.
\item Our observations reveal a regime shift in upper--ocean turbulence linked to sea ice retreat, exposing a critical ocean-ice feedback unresolved in climate models.
\item High-resolution simulations capture the regime shift but overestimate enstrophy, underscoring the urgent need for more observations to constrain model physics.

\end{keypoints}

\begin{abstract}  
Quantifying kinetic energy (KE) and enstrophy transfer, mixing, and dissipation in the Arctic Ocean is key to understanding polar ocean dynamics, which are critical components of the global climate system. However, in ice-covered regions, limited eddy-resolving observations challenge characterizing KE and enstrophy transfer across scales. Here, we use satellite-derived sea ice floe rotation rates to infer the surface ocean enstrophy spectra in the marginal ice zone. Employing a coarse-graining approach, we treat each floe as a local spatial filter. The method is validated with idealized sea ice-ocean simulations and applied to floe observations in the Beaufort Gyre. Our results reveal steepened spectral slopes at low sea ice concentrations, indicating enhanced mesoscale activity during the spring-to-summer transition. High-resolution simulations support these findings but overestimate enstrophy, highlighting a denser array of observations. Our two-dimensional spectral estimates are the first of their kind, providing a scalable approach for mapping Arctic Ocean characteristics.
\end{abstract}

\section*{Plain Language Summary}
Measuring energy transfer in the Arctic Ocean is essential for understanding how polar oceans work, as it influences large-scale circulation and global climate. However, studying ice-covered regions is challenging due to limited under-ice observations. In this study, we use the rotation of sea ice plates (floes) to estimate how much ocean enstrophy, a measure of turbulence and swirling motion, is present at different scales of surface oceanic motion. Each floe is treated as a local sampling area, and the method is validated with computer simulations of sea ice and ocean flow. We then apply it to satellite observations of ice floes available during spring and summer, finding that the rate of change in enstrophy from large to small scales becomes steeper when sea ice concentration is low. This trend suggests stronger large-scale activity and weaker small-scale motion compared to periods in which there is high sea ice concentration. Our approach provides a robust framework, leveraging publicly available satellite data to map Arctic ocean dynamics over both space and time.

\section{Introduction}

Kinetic energy (KE) and enstrophy are fundamental diagnostics of the ocean energy budget, governing mixing processes that drive the large-scale circulation, redistribute energy, transport heat and salt, and influence climate variability. A key aspect of both quantities is their spectral distribution, which characterizes the dominant spatial scales and the mechanisms for energy transfer, mixing, and dissipation. While KE spectra have been extensively studied, enstrophy spectra have received comparatively less attention, despite their critical role in characterizing forward cascades in two-dimensional turbulence \cite{Tennekes1972} and their close connection to KE spectra \cite{Janosi2022,Barkan2024}; in geostrophically dominated regimes, their slopes differ by exactly two.


In ice-covered regions, sea ice modulates upper-ocean energetics via drag, influencing the seasonal variability of KE spectra \cite{Timmermans2012,Zhao2018,Wang2020,Cassianides2021,Appen2022,Manucharyan2022d,Cassianides2023,Liu2024,Muller2024,Martinez2024}. Due to the lack of 2D surface velocity observations, 1D along-track profiles from ice-tethered profilers (ITPs) and surface drifters (in open water) have been used to compute potential density variance spectra, a proxy for ocean potential energy (PE) linked to KE \cite{Timmermans2012,Timmermans2013,Mensa2018,Cassianides2023}. 
These studies found that submesoscale activity tends to increase in summer \cite{Cassianides2023} and decrease in winter \cite{Timmermans2013}, in contrast to observations from the open ocean \cite{Callies2015,Khatri2018,Storer2022}, the Southern Ocean \cite{Thompson2016}, and the Antarctic marginal ice zone (MIZ) \cite{Biddle2020}, where submesoscale activity typically intensifies in winter. These inconsistencies highlight the need for 2D observations, as ITPs provide limited spatial coverage and are insufficient for reconstructing full KE or enstrophy spectra.

Numerical simulations have emerged as a critical tool to address the gaps left by sparse observations of under-ice ocean turbulence \cite{Horvat2016,Manucharyan2017,Mensa2017,Horvat2018,Armitage2020,Wang2020,Gupta2022,Manucharyan2022b,Manucharyan2022d,Shrestha2022,Brenner2023,Gupta2024,Liu2024,Martinez2024,Muller2024,Mason2025}.
These simulation-based studies show that mesoscale eddies, generated by interior baroclinic instabilities, lose energy in the ice-ocean boundary layer \cite{Manucharyan2022b}, leading to suppressed mesoscale motions in winter \cite{Martinez2024}. In addition, sea ice growth and melt drive mixed-layer instabilities and eddy formation \cite{Manucharyan2022d,Liu2024}, contributing to the seasonal variability of upper-ocean KE spectra. Despite the small spatial scales involved in these processes, there can be large-scale consequences: as Arctic sea ice declines, KE is projected to increase--especially in the MIZ--potentially accelerating melt and amplifying the ocean--atmosphere feedback \cite{Armitage2020,Li2024}. While simulations capture eddy generation mechanisms under sea ice, they still need further validation and refinement, as most models cannot fully resolve mixed-layer eddies and often rely on parameterizations calibrated using ice-free conditions. 

Recent studies have proposed inferring ocean turbulence from the kinematics of ice floes observed via satellite remote sensing. In particular, ice floe rotation about its center of mass has emerged as a promising proxy for upper-ocean vorticity \cite{Lopez2021,Manucharyan2022c,Manucharyan2022d,Brenner2023,Kim2025}. While atmospheric winds primarily drive floe translation, their large spatial scales ($\sim O(10^3)$ km) have minor effects on rotation. Instead, floe rotation is governed by local ocean vorticity \cite{Manucharyan2022c,Kim2025}, which exhibits longer coherent timescales ($\sim O(1)$ days). These findings highlight the potential of floe rotation rates to infer upper-ocean enstrophy.

In the present study, we use spinning sea ice floes as natural tracers to directly estimate surface ocean enstrophy spectra underneath sea ice. To compute enstrophy spectra, we treat each floe as a local spatial filter, analogous to standard coarse-graining kernels \cite{Aluie2018,Storer2022,Buzzicotti2023,Liu2024,Khatri2024}, and validate this approach using idealized sea ice-ocean simulations. Applying the method to observed ice floe rotation data in the Beaufort Gyre (BG) MIZ, we infer enstrophy spectra across different SIC regimes and evaluate them in contrast to spectra derived from high-resolution ocean–sea ice simulations. This framework provides a paradigm-shifting approach for characterizing ocean turbulence beneath sea ice and assessing the fidelity of state-of-the-art high-resolution ocean models.

\section{Methods and Data} \label{sec:methods}

\subsection{Sea ice floe rotation measurements from optical satellite imagery}

\begin{figure}
  \centerline{\includegraphics[width=\linewidth]{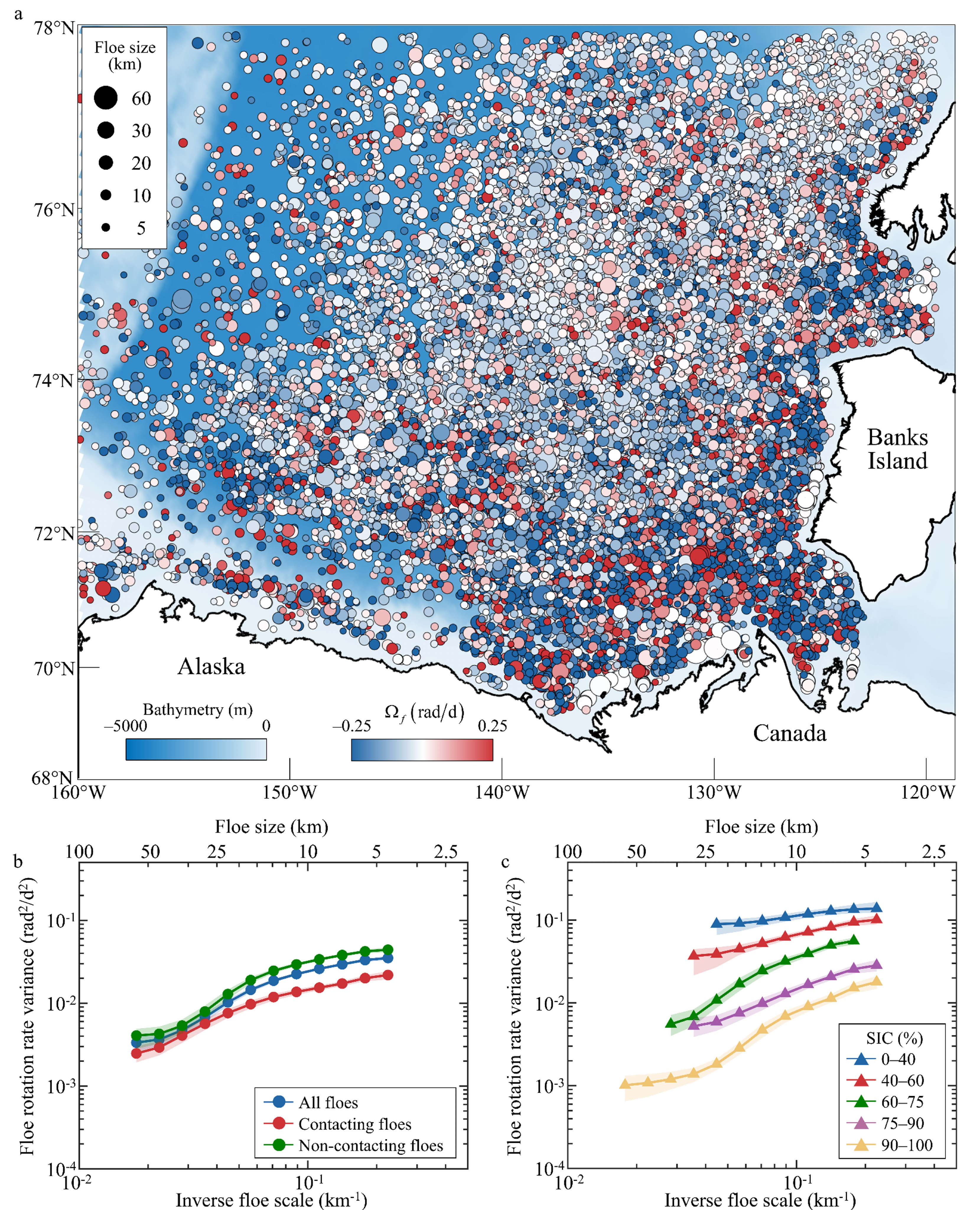}}
  \caption{Geographic location and physical properties of observed ice floes in the BG. (a) Geospatial distribution of ice floes, in which the size of each circular marker scales with floe size, and the colors indicate ice floe rotation rates. Bathymetry is shown using the International Bathymetric Chart of the Arctic Ocean. (b,c) The variance of ice floe rotation rates as a function of inverse floe scale, computed from binned floe sizes, is shown for (b) different floe-floe contact scenarios and (c) SIC ranges for non-contacting floes. Shading shows 95\% bootstrap confidence intervals. Corresponding floe sizes are shown on the top $x$-axis.} \label{Fig:IFT_data}
\end{figure}

We used the recently developed ice floe tracking (IFT) algorithm by \citeA{Lopez2019} to retrieve rotation rates of sea ice floes in the BG (Figure \ref{Fig:IFT_data}a). We employed Moderate Resolution Imaging Spectroradiometer (MODIS) Level 1B optical imagery from the Aqua and Terra satellites. Daily Corrected Reflectance True Color and False Color images (250 m) were employed to distinguish sea ice from clouds and open water, extract ice floe features, and compute rotation rates about the center of mass of each floe, $\Omega_f$. The ice floe dataset (4--75 km in floe size, defined by the square root of the area) spans the spring and summer months (May--August) from 2003 to 2020. We then grouped floes by (i) the presence of floe-floe contact, based on geometric and kinematic criteria, and (ii) the SIC computed at the floe centroid, determined from the NOAA/NSIDC Climate Data Record of Passive Microwave SIC (25 km $\times$ 25 km resolution). We compute the variance of observed ice floe rotation rates, $\Omega_f'^2$, binned by floe size (Figures \ref{Fig:IFT_data}b and \ref{Fig:IFT_data}c): $\langle \Omega_f'^2 (k_f) \rangle = \langle (\Omega_f - \langle \Omega_f \rangle)^2 \rangle,$ where $k_f=1/L_f$ is the inverse floe scale, $\langle\rangle$ denotes bin-averaging by floe size $L_f$. Further details on the algorithm and floe grouping are provided in Supporting Information S1 and in \citeA{Lopez2019}, \citeA{Lopez2021}, and \citeA{Watkins2024}.

\subsection{Coarse-graining approach to infer ocean enstrophy spectra} \label{subsec:cg}

We use a coarse-graining (CG) framework to infer upper-ocean enstrophy spectra from ice floe rotation rates. Similar techniques have been applied to derive KE spectra from satellite-derived surface ocean velocity fields \cite{Storer2022} and high-resolution global ocean model simulations \cite{Aluie2018,Storer2022,Liu2024,Muller2024,Khatri2024}. We extend this framework to vorticity following the velocity-based CG formulation of \citeA{Sadek2018} (see Supporting Information S2). Considering the rotational component of the ocean field $\omega_o(\mathbf{r})$, the coarse-grained vorticity at filter size $l$ is defined as:
\begin{equation} \label{Eq:w_bar}
    \overline{\omega}_{o,l}(\mathbf{x}) = \int_A G_l(\mathbf{x}-\mathbf{r})\omega_o(\mathbf{r})d^2\mathbf{r},
\end{equation}
where $G_l$ is a normalized spatial filter (or kernel; top-hat), $\mathbf{x}$ is the evaluation point in space, $\mathbf{r}$ is the integration variable, and $d^2$ denotes integration over a two-dimensional domain $A$. The filtered enstrophy spectrum is then:
\begin{equation} \label{Eq:CG_spec}
    \overline{E}_{\omega}(k_l) = \frac{d}{dk_l} \left( \sum_{\mathbf{x}} \frac{1}{2} \overline{\omega}_{o,l}^2  \left( \mathbf{x}\right)  \right),
\end{equation}
where $k_l = L/l$ is the dimensionless wavenumber, with $L$ being a characteristic length scale. We demonstrate that this approach reproduces the slope of the Fourier enstrophy spectrum when it lies between 0 and 3. We then apply the vorticity-based CG framework to ice floes, which act as spatial filters by locally sampling ocean vorticity. Assuming the floes behave as rigid bodies over $\mathcal{O}(1)$-day timescales, they approximate first-order kernels in Equation \ref{Eq:w_bar}, where $\mathbf{r}$ spans the floe surface relative to the centroid $\mathbf{x}$. We relate observed ice floe rotation rates to surface ocean vorticity via $\omega_f \approx 2\Omega_f$ \cite{Manucharyan2022c,Kim2025}, yielding $\overline{\omega}_{o,l} \approx 2\Omega_f$, allowing estimation of the filtered enstrophy spectrum $\overline{E}_\omega(k_l)$.

\subsection{Floe rotation in a generalized quasi-geostrophic (QG) model of ocean turbulence} \label{subsubsec:AQG}

\begin{figure}
  \centerline{\includegraphics[width=\linewidth]{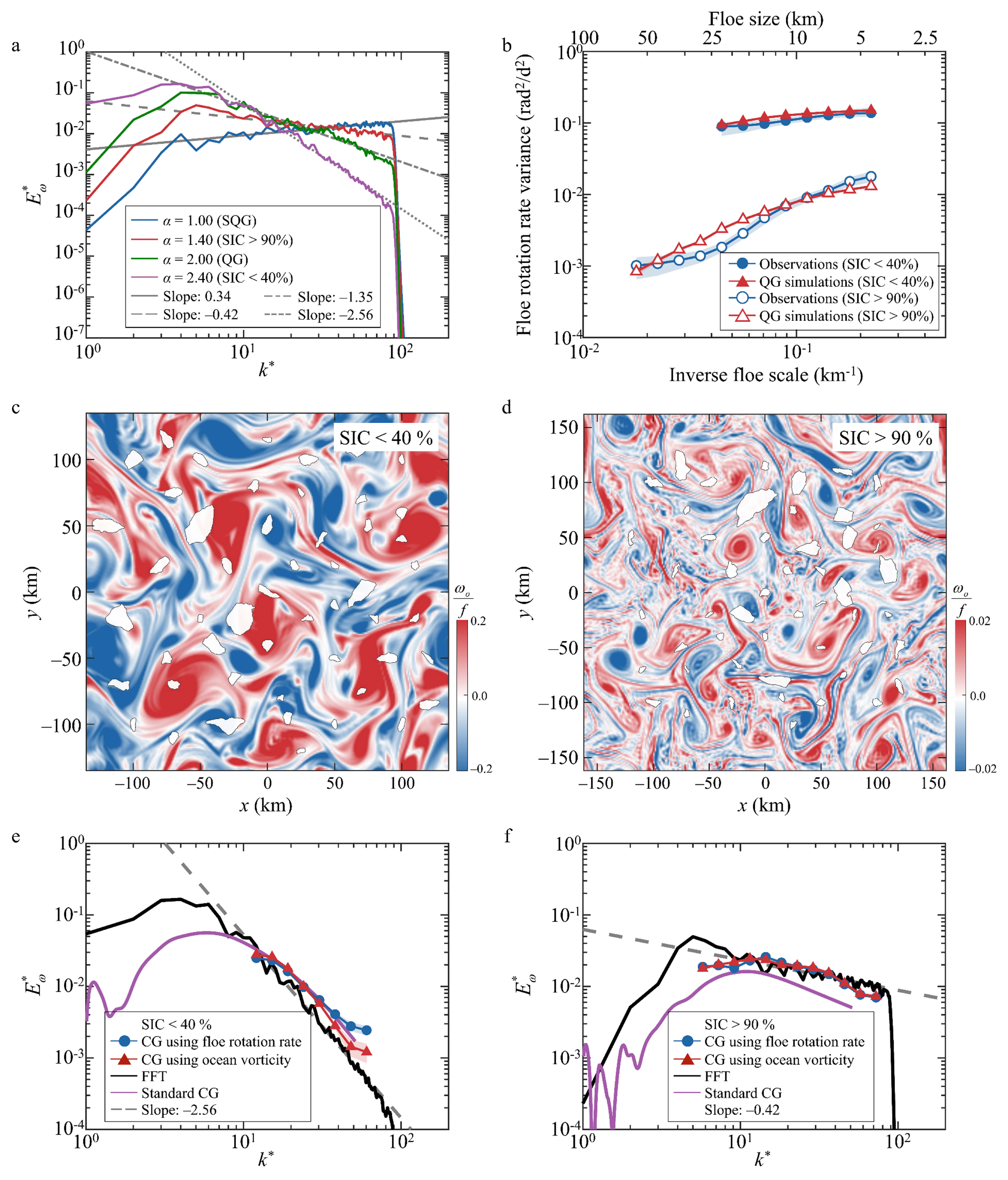}}
  \caption{Validation of the CG approach using idealized sea ice-QG ocean turbulence simulations. (a) Normalized enstrophy spectra for different model parameters. (b) Variance of ice floe rotation rates as a function of inverse floe scale for low ($<$ 40 $\%$) and high ($>$ 90 $\%$) SIC. Shading shows 95\% bootstrap confidence intervals. The top $x$-axis shows corresponding floe sizes. (c,d) Non-dimensionalized vorticity fields overlaid with simulated floes (white objects) under (c) low and (d) high SIC. (e,f) Normalized enstrophy spectra from CG applied to ice floe rotation rates and averaged ocean vorticity beneath the floes, compared with Fourier Transform and standard CG, for (e) low and (f) high SIC. Shaded bands represent interquartile ranges. Fitted lines indicate slopes over the inertial range.} \label{Fig:IFT_method}
\end{figure}

Given the lack of co-located rotational and spectral observations, we use a generalized QG ocean model, known as the $\alpha$QG model \cite{Pierrehumbert1994,Held1995,Foussard2017}, to assess how accurately inferred enstrophy spectra reproduce those directly derived from the modeled velocity field (Figure \ref{Fig:IFT_method}). The model produces flow fields spanning a wide range of spectral slopes and associated energy distributions (Figure \ref{Fig:IFT_method}a), where the evolution of a conserved scalar $q$ in a velocity field $\mathbf{u}$ is governed by: 
\begin{equation} \label{Eq:AQG_ori}
    \frac{\partial q}{\partial t} + (\mathbf{u} \cdot \nabla)q = 0, \qquad \hat{q} = -k^\alpha \hat{\psi},
\end{equation}
where $t$ is time, $\psi$ is the streamfunction, the hat symbol denotes the Fourier transform, and $k$ is the wavenumber. Note that an exponential filter is applied to provide numerical dissipation at small spatial scales \cite{Lacasce1996}. The turbulence regime and resulting enstrophy spectral slope $\beta$ are determined by $\beta=(4\alpha-5)/3$ for $0 < \alpha < 2$ \cite{Foussard2017}; for example, $\alpha=1$ corresponds to surface QG, while $\alpha=2$ corresponds to interior QG.

Solid-plate floes with realistic shapes derived from observed floe tracking data are randomly seeded over simulated QG velocity fields (Figures \ref{Fig:IFT_method}c and \ref{Fig:IFT_method}d). They are set into motion using Subzero, a discrete-element sea ice model \cite{Manucharyan2022a,Montemuro2023}. The $\alpha$QG equations are dimensionalized using the parameter $\alpha$, a reference velocity $U_{\mathrm{ref}}$, and a reference length scale $L_{\mathrm{ref}}$. Full details on the models are provided in Supporting Information S3.

\subsection{High-resolution global ocean and sea ice simulation}

We compare satellite-inferred ocean enstrophy spectra with model-based estimates from the MIT general circulation model configured on a Lat-Lon-Cap (LLC) grid, a quasi-uniform global grid that combines latitude-longitude and polar cap tiles, with a nominal horizontal resolution of 1/48$^\circ$ ($\sim$0.8 km in the Arctic Ocean), known as LLC4320 \cite{Gallmeier2023}. We use daily-averaged ocean velocity fields at 20 m depth, representing surface conditions within the typical mixed layer depth ($\sim O(10)$ m) \cite{Toole2010,Timmermans2012}. Our analysis focuses on the Beaufort Sea during the spring-to-summer transition (May 1 to August 31), where the first Rossby radius of deformation is approximately 11--13 km \cite{Nurser2014,Zhao2014} (see Supporting Information S4 for model details).

\section{Results} \label{sec:results}

\subsection{Validating ice floes as effective CG filters for estimating the enstrophy spectrum} \label{subsec:val_cg}

We validate the floe-based CG approach using sea ice–QG ocean simulations. First, we compare the variance of simulated and observed ice floe rotation rates (Figure \ref{Fig:IFT_method}). The $\alpha$QG model parameters ($\alpha$, $L_{\mathrm{ref}}$, and $U_{\mathrm{ref}}$) are tuned to match observed rotational variances for floe sizes in low ($< 40\%$) and high ($> 90\%$) SIC regimes (Figure \ref{Fig:IFT_method}b). Agreement with observations is then quantified using a loss function: $\mathcal{L}_\Omega = |\langle\Omega_f'^2\rangle_{\mathrm{sim}} - \langle\Omega_f'^2\rangle_{\mathrm{obs}}| / \langle\Omega_f'^2\rangle_{\mathrm{obs}}$ \cite{Manucharyan2022c}, where the subscripts ``sim" and ``obs" refer to simulation and observation values, respectively. Full details of the tuning procedure are provided in Supporting Information S3. The tuned QG flow fields show distinct characteristics between SIC regimes: under low SIC, the flow exhibits high vorticity and larger-scale coherent structures (Figure \ref{Fig:IFT_method}c), resulting in a steeper enstrophy spectrum slope (Figure \ref{Fig:IFT_method}a). In contrast, the high SIC case is dominated by small-scale structures with lower vorticity, producing a shallower slope (Figure \ref{Fig:IFT_method}d).

Second, we compare CG-inferred enstrophy spectra with Fourier and standard CG-based spectra across low and high SIC conditions, finding strong agreement between the methods (Figures \ref{Fig:IFT_method}e and \ref{Fig:IFT_method}f). Here, spectra obtained from the Fourier transform of the tuned QG flow fields are used as the ground truth, while standard CG enstrophy spectra are computed by applying a top-hat filter to the same flow fields. In both SIC regimes, the floe-based CG spectra closely match the Fourier-derived slopes and show good agreement with the standard CG results. We also compute enstrophy spectra using the averaged ocean vorticity beneath each floe. Both approaches yield similar spectral slopes, supporting the idea of using daily ice floe rotation observations to infer ocean enstrophy spectra. Further discussion of the effects of wind forcing and floe shape is provided in Supporting Information S5. In the following section, we extend this approach to investigate enstrophy spectra under different floe-floe contact conditions and across SIC regimes in observations of the BG.


\subsection{Ocean enstrophy spectra in the BG MIZ} \label{subsec:Ens_spec_BG}

\begin{figure}
  \centerline{\includegraphics[width=\linewidth]{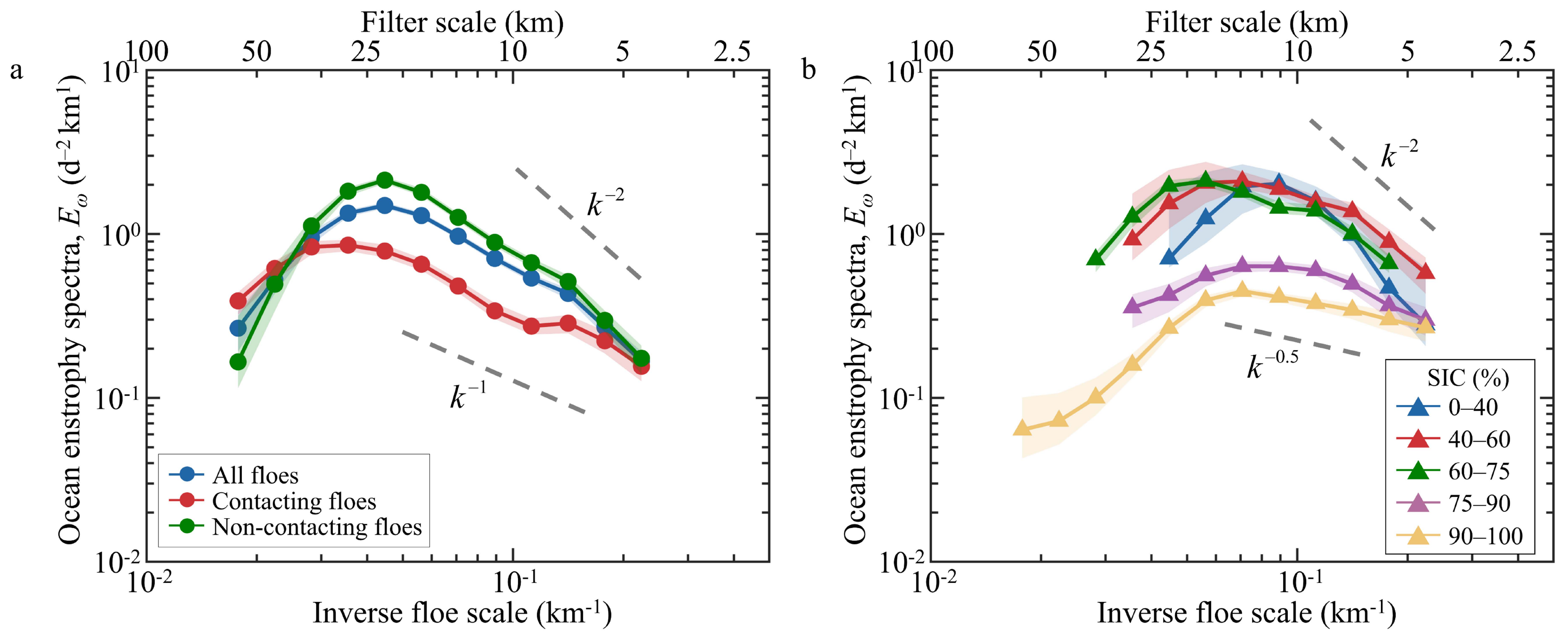}}
  \caption{Inferred ocean enstrophy spectra as a function of inverse floe scale for (a) different floe-floe contact conditions and (b) non-contacting floes under different SIC ranges. Shaded bands represent interquartile ranges. Dashed gray lines indicate reference slopes for comparison.} \label{Fig:IFT_spectra}
\end{figure}

We analyze ocean enstrophy spectra inferred from observed ice floe rotation rates under different floe-floe contact conditions and SIC regimes (Figure \ref{Fig:IFT_spectra}). As described in Section \ref{subsec:val_cg}, rotation rate variance is used to estimate enstrophy spectra via the CG formulation (Equation \ref{Eq:CG_spec}).

For all contact types, the spectra peak at $k_f = 0.03-0.04$ km$^{-1}$ (Figure \ref{Fig:IFT_spectra}a), corresponding to $L_f = 25-33$ km, which is significantly smaller than the spectral peaks of mid-latitude altimetry-derived spectra \cite{Khatri2018,Storer2022}. In both regions, however, the spectra typically peak at scales larger than the first baroclinic Rossby radius of deformation (11$-$13 km in the BG and 20$-$50 km at mid-latitudes) \cite{Khatri2018,Storer2022,Liu2024}.

Spectral slopes are computed over $k_f = 0.05-0.2$ km$^{-1}$ ($L_f = 5-20$ km), excluding the peak to avoid slope bias. This range is consistent with prior scaling analyses in the Finite Element Sea Ice-Ocean Model (FESOM2) simulations \cite{Liu2024}. For non-contacting floes, the enstrophy spectrum is consistent with strong geostrophy dynamics, with a slope steeper than the commonly reported value of $-3$ for interior QG turbulence \cite{Khatri2018}. Finally, contacting floes exhibit a shallower spectrum slope, while the combined spectrum falls in between.

Enstrophy spectra also vary with SIC conditions (Figure \ref{Fig:IFT_spectra}b). All cases peak near $k_f = 0.05-0.08$ km$^{-1}$ ($L_f = 13-20$ km), with peak wavenumbers shifting from $k_f = 0.08$ to $0.05$ km$^{-1}$ as SIC decreases. Spectra are truncated at small $k$ given the limited number of large floes in the dataset. For SIC $<60\%$, spectral values are similar, whereas they are significantly reduced at higher SIC, indicating a decrease in enstrophy content. As SIC increases, spectral slopes become shallower, transitioning from approximately $k^{-2}$ to $k^{-0.5}$ in enstrophy (equivalent to $k^{-4}$ to $k^{-2.5}$ in KE). This suggests enhanced submesoscale activity relative to mesoscale motions, despite the overall reduction in enstrophy. We analyze the derived slopes and enstrophy content in the following section.

\subsection{The enstrophy spectra and their seasonal variability in the BG} \label{subsec:Ens_slope_BG}

\begin{figure}
  \centerline{\includegraphics[width=0.8\linewidth]{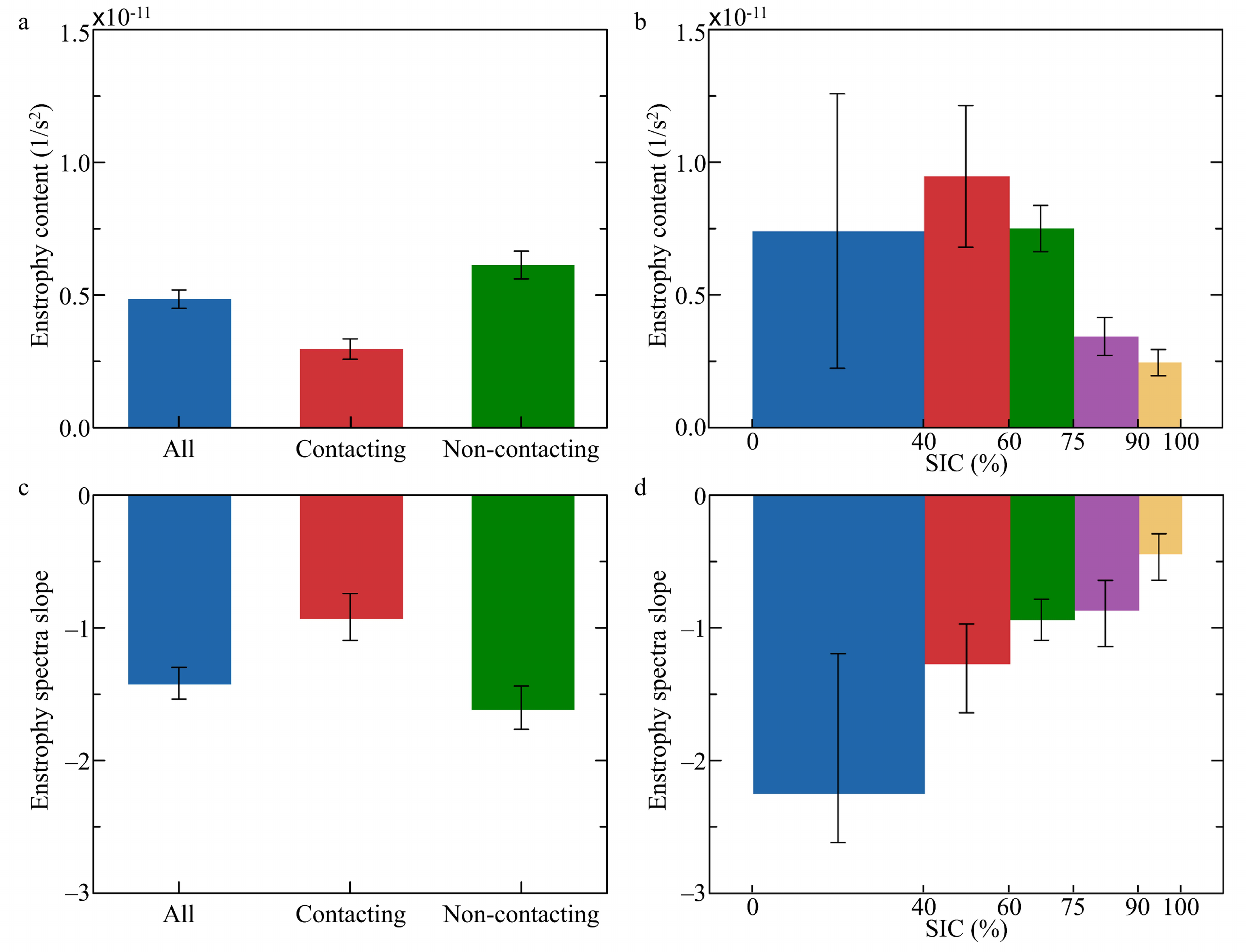}}
  \caption{Estimated ocean enstrophy content and spectral slopes derived from ice floe rotation rates for different floe-floe contact conditions and SIC regimes. (a,b) Enstrophy content and (c,d) spectral slopes computed for (a,c) different contact conditions and (b,d) non-contacting floes under different SIC bins.} \label{Fig:IFT_bar}
\end{figure}
 
We analyze the ocean enstrophy content and spectral slopes inferred from ice floe rotation rates for different floe–floe contact conditions and SIC regimes (Figure \ref{Fig:IFT_bar}). Enstrophy content is computed over $L_f = 4–28$ km, and slopes are based on the best fits to the spectra in Figure \ref{Fig:IFT_spectra}. It is worth noting that mean SIC drops markedly from $\sim 85 \%$ in early June to $\sim 40 \%$ by late August (2003–2020 average; NOAA/NSIDC \cite{Meier2021}; see Supporting Information S6), reflecting a strong seasonality. This seasonal SIC decline is closely tied to shifts in upper-ocean energetics \cite{Manucharyan2022d,Martinez2024}.

Independent of SIC, non-contacting floes show the highest enstrophy content in the 4$-$28 km band due to less rotational damping, which reduces $\Omega_f$ in contacting cases (Figure \ref{Fig:IFT_bar}a). Corresponding spectral slopes are $-$1.6 (non-contacting), $-$0.9 (contacting), and $-$1.4 (all floes) (Figure \ref{Fig:IFT_bar}c). Contact-induced noise flattens slopes, similar to altimetry-derived KE spectra affected by measurement uncertainty \cite{Xu2012}. The implied KE spectral slope from non-contacting floes ($\sim–3.6$) lies within the range observed at mid-latitudes over 100$-$200 km scales \cite{Callies2015,Khatri2018,Storer2022}; for instance, altimetry-derived spectra often exceed –3 \cite{Khatri2018}.

Both enstrophy content and slopes are strongly SIC-dependent. As SIC increases, enstrophy content declines, consistent with suppressed eddy KE \cite{Wang2020,Liu2024,Manucharyan2022b} (Figure \ref{Fig:IFT_bar}b). Enstrophy spectra also flatten: slopes transition from $-$2.25 (SIC $\leq40\%$) to $-$0.44 (SIC $\geq90\%$), reflecting a shift from interior-QG to submesoscale-dominated regimes (Figure \ref{Fig:IFT_bar}d), despite the lower enstrophy content at higher SIC.

\subsection{Comparison with high-resolution ocean simulations} \label{subsec:LLC}

\begin{figure}
  \centerline{\includegraphics[width=\linewidth]{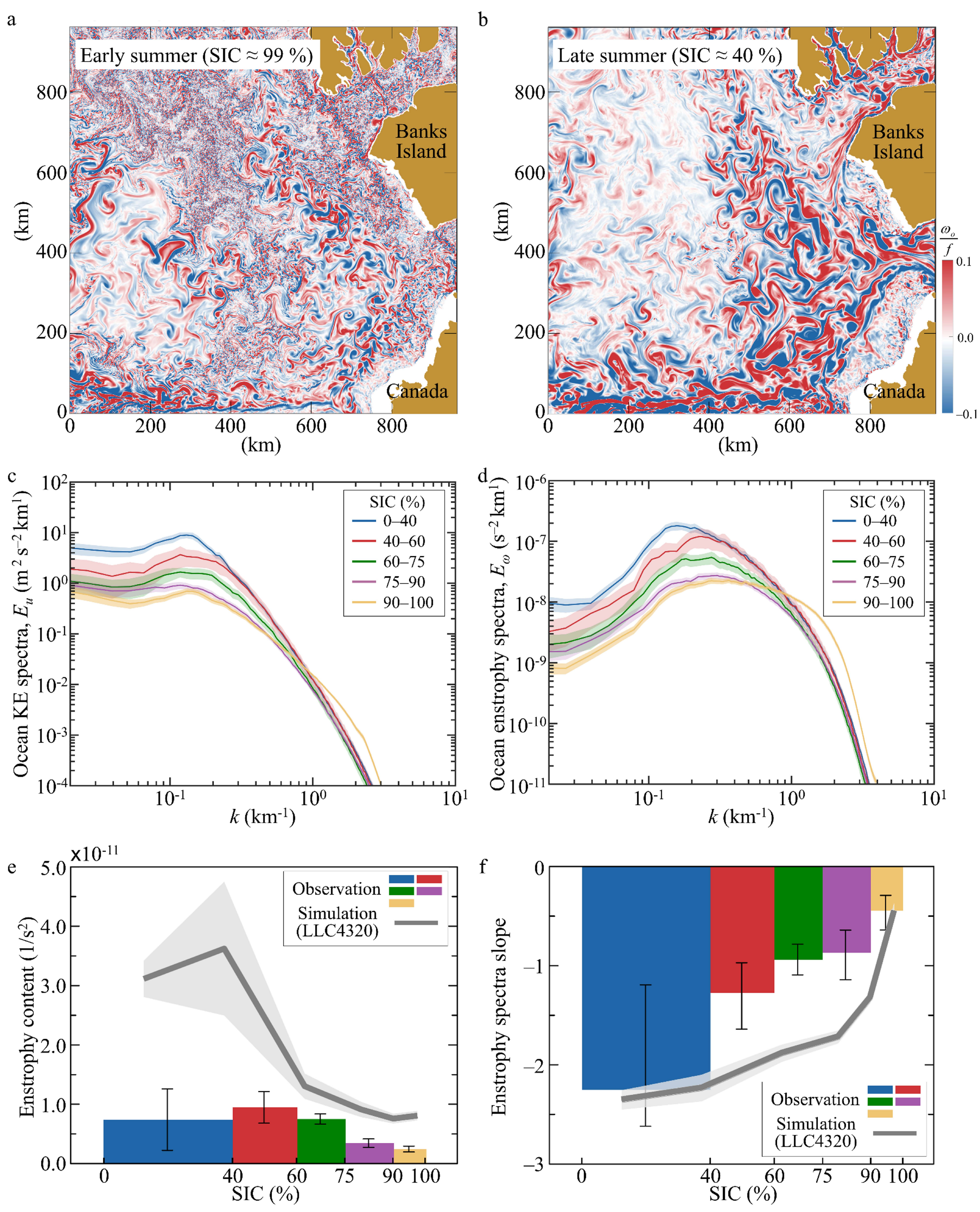}}
  \caption{Ocean eddy KE and enstrophy spectra from the LLC4320 simulation data for different SIC conditions between May and August 2012. Daily-averaged vorticity fields at 20 m depth for (a) May 5th and (b) August 23rd. SIC-binned (c) KE and (d) enstrophy spectra averaged over sub-regions in the BG. SIC-binned (e) enstrophy content and (f) enstrophy spectral slopes are compared with those derived from observed ice floe rotation rates.} \label{Fig:LLC}
\end{figure}

We compare the ocean enstrophy content and spectra inferred from ice floe observations with those derived from the high-resolution global ocean simulation data, LLC4320 (Figure \ref{Fig:LLC}). Simulated vorticity fields at 20-m depth show distinct structures under different SIC conditions (Figures \ref{Fig:LLC}a and \ref{Fig:LLC}b). In early summer (high SIC, $\approx 99\%$), fine vortices and filaments prevail across the study region, whereas in late summer (low SIC, $\approx 40\%$), stronger and larger vortices emerge, concentrated along the coastlines. These trends are consistent with the higher enstrophy content and steeper spectra observed under low SIC in the floe-based analysis (Figures \ref{Fig:IFT_bar}b and \ref{Fig:IFT_bar}d).

We compute KE and enstrophy spectra in the BG by applying the Fourier transform to velocity and vorticity fields in 11 sub-regions (240 km each). To isolate small-scale variability, 5-day averaged flow fields at 20 m depth are subtracted from each instantaneous flow field. We calculate spectral slopes over $k = 0.3$–$1.0$ km$^{-1}$ (6$-$20 km scales), matching the floe-based spectra scaling range. Enstrophy content in the 4$-$28 km band is estimated using a standard CG method and similarly averaged. The resulting spectra show strong SIC dependence (Figures \ref{Fig:LLC}c and \ref{Fig:LLC}d). Across all SIC levels, spectral peaks occur at $k=0.1-0.2$ km$^{-1}$ (30$-$60 km scales), comparable to peaks inferred from non-contacting floes (25$-$33 km scales). Peak amplitudes decline and spectra become shallower with increasing SIC, indicating reduced overall enstrophy content and relatively enhanced submesoscale activity during the early summer. 

The LLC-derived enstrophy content is about three times higher than the floe-based observational estimates, while both peak near SIC $\sim 40 \%$ before significantly decreasing at higher SICs (Figure \ref{Fig:LLC}e). In contrast, LLC- and floe-based spectral slopes are comparable at low ($<40\%$) and high ($>90\%$) SICs, whereas LLC-derived spectra are steeper at intermediate SIC levels (Figure \ref{Fig:LLC}f). It is important to note that the KE--enstrophy slope relationship in the LLC output does generally follow geostrophic scaling, with minor deviations ($\sim 0.2$) likely due to ageostrophic processes near the first Rossby deformation radius (see Supporting Information S4). This provides strong evidence for using enstrophy spectra as a proxy for estimating KE spectra, particularly in geostrophically dominated regimes.


\section{Discussion and conclusion} \label{sec:discussion}

We used sea ice floe satellite observations to provide the first direct observational estimates of surface ocean enstrophy content and spectra in the Arctic by adopting a CG framework. Treating ice floes as spatial filters that capture locally integrated ocean vorticity, we conducted a statistical averaging of floe rotation rates to compute the enstrophy content and spectra. By applying this new method to BG ice floe observations, we reveal a strong dependence of enstrophy content and spectral slopes on SIC. As SIC declines sharply from June to August, enstrophy content in the 4--28 km band increases and spectral slopes steepen, indicating suppressed submesoscale activity relative to mesoscale motions, despite enhanced overall enstrophy. While these SIC-dependent behaviors are also reproduced in high-resolution ocean simulations (LLC4320), LLC-derived enstrophy content is consistently higher than floe-based observational estimates, and deviations in spectral slopes emerge at intermediate SIC levels. The suppression of small-scale eddy generation, likely caused by poorly resolved mixed-layer instabilities, is a probable source of these discrepancies, and thus capturing the appropriate spectral contents and slopes requires better constrained high-resolution simulations.

The observed SIC dependence in surface ocean enstrophy spectra during the spring-to-summer transition is consistent with near-surface KE spectra in the BG from high-resolution simulations \cite{Liu2024,Martinez2024}. For example, surface ocean KE spectral slopes are $-5/3$ in August--January and $-3$ in February--July in FESOM2 simulations \cite{Liu2024} and steepen beyond $-3$ during June--July and become shallower than $-5/3$ during winter and spring in high-resolution sea ice-ocean simulations \cite{Martinez2024}. This seasonal trend is attributed to changes in mixed-layer depths (MLD), where deep winter MLDs enhance baroclinic instability, whereas shallow summer MLDs promote dissipation \cite{Martinez2024}. Similar variability is evident in the BG MLD observations, which show depths of approximately 30 m in March--April and less than 9 m in summer \cite{Peralta2015}.

Our floe-based analysis and findings are validated through simulations, yet the influence of additional oceanic processes (internal waves, tides, inertial oscillations) and sea ice properties (floe-floe interactions, shapes, inertia) may affect spectral shape and introduce uncertainties in the retrieval. As sea ice melts and damping weakens during the summer, internal waves---known to flatten KE spectra at small scales \cite{Callies2015}---may have stronger influence \cite{Martini2014}. Tides can broaden spectral peaks \cite{Callies2013}, though they are generally weak in Arctic basins \cite{Martini2014}, while inertial oscillations, with sub-daily frequencies in the Arctic \cite{Gimbert2012}, may contribute to seasonal spectral variability \cite{Martini2014}. During the floe retrieval process, these oceanic motions can drive horizontal floe excursions, primarily influencing floe translation \cite{Martini2014}. In addition, floe–floe contact introduces internal stress \cite{Gimbert2012} and heterogeneity in floe thickness and surface roughness \cite{Feltham2008}, which add uncertainty to the relationship between floe rotation and ocean vorticity. Further integration of high-resolution observations into our framework will enable quantification of these effects and strengthen the robustness of floe-based retrievals. Our method moves beyond the limitations of numerical models to provide the first direct, two-dimensional observational measurements of surface ocean enstrophy spectra beneath sea ice. Despite the complexities discussed, this work opens a unique pathway for mapping and characterizing Arctic Ocean turbulence.
\section{Open Research} \label{sec:open_research}
IFT datasets used for this analysis are available at \url{https://doi.org/10.5281/zenodo.7996638}. The MODIS imagery was downloaded from NASA Earthdata Worldview (\url{https://worldview.earthdata.nasa.gov}), part of NASA Earth Observing System Data and Information System (EOSDIS). The International Bathymetric Chart of the Arctic Ocean (IBCAO) bathymetric data at 400 m by 400 m resolution was downloaded from \url{https://www.gebco.net/data-products/gridded-bathymetry-data/arctic-ocean/}. The codes associated with sea ice-QG ocean turbulence simulations are available at \url{https://github.com/SeaIce-Math/SubZero}. The code and data used to generate the figures are available at \url{https://zenodo.org/records/17083317}.






\acknowledgments
M.K., D.M.W, and M.M.W were supported by the ONR Arctic Program (N00014-20-1-2753, N00014-22-1-2741, and N00014-22-1-2722) and by the NASA Science of Terra, Aqua, and Suomi-NPP Program (80NSSC22K0620, 20-TASNPP20-0202). M.K., M.M.W, and G.E.M were also supported by the ONR Multidisciplinary University Research Initiatives Program (N00014-23-1-2014) and the ONR Young Investigator Program (N00014-24-1-2283). G.E.M was supported by the NSF CAREER award (2338233). 
 

\bibliography{Main}
%
%




%
%
%
%
%


\end{document}